\documentclass[jkps,preprint,fleqn,showpacs,showkeys]{revtex4}
\usepackage{graphicx}
\usepackage{amssymb,amsmath,bm}

\begin{document}

\title{Entanglement Measure of the planar-transverse classical light field }

\author{Sun-Hyun Youn\footnote{email: sunyoun@chonnam.ac.kr, fax: +82-62-530-3369}}

\address{Department of Physics, Chonnam National University, Gwangju 500-757, Korea}

\begin{abstract}

Classical light fields are considered physical examples of
nonquantum entanglement\cite{Eberly2011}. We apply concurrence and
Schmidt approach to evaluate the degree of entanglement for a
generalized polarization state that Qian and Eberly suggested, and
we obtained an analytic form of the general entanglement state
for planar polarization states of classical light.



\end{abstract}


\maketitle

\section{Introduction}

Entanglement was considered a purely quantum mechanical property
until  Eberly suggested physical examples of nonquantum
entanglement\cite{Eberly2011}. They identified the polarization of
classical light fields as a physical example of nonquantum
entanglement.
 After Stokes defined the polarization state \cite{Stokes}, the
formulation of the polarization state has continuously evolved and
is now well established in terms of field correlation functions
\cite{Wolf}. The familiar measures of polarization come from the
paraxial field approximation. But,the traditional picture of
polarization must be reconsidered   for fully three-dimensional
fields such as highly nonparaxial fields\cite{Ash,
Pohl,Moore,Brosseau}

 Those studies extended  the electric field from planar-transverse to
nonplanar such as
\begin{eqnarray}
  \vec{E}   = {\vec x } E_{x} + {\vec y} E_{y}  \Rightarrow  {\vec x } E_{x} + {\vec y} E_{y} +{\vec z} E_{z},     \label{eqEfield}
\end{eqnarray}
and employed two independent vector spaces that are entangled to
realize ${\vec E}$ in Eq. \ref{eqEfield}. Entanglement is a
technical term indicated by ${\vec E}$ in Eq. \ref{eqEfield}. It
is a tensor product of "lab space" unit vectors, such as ${\vec
x}$ and ${\vec y}$, and functions $E_{x}$ and $E_{y}$, which are
vectors in a statistical "function space" of continuous normed
functions.

 In a mathematical sense, determining the degree of
polarization is the same as the determining degree of
factorization of two spaces.
 In this work, we obtained an complete form of the general entanglement state
for planar polarization states of classical light.

\section{Schmidt value and concurrence}
\subsection{Schmidt value}

 Two independent vector spaces are introduced to express the electric
field. ${\vec{E}}$ is a tensor product of "lab space" unit
vectors, such as ${\vec x}$ and ${\vec y}$, and functions $E_x$
and $E_y$, which are vectors in a statistical "function space" of
continuous normed functions\cite{Eberly2011}.  With intensity $I =
<E_x | E_x >+ <E_y | E_y > $ factored out,

\begin{eqnarray}
 | \vec{E} >/ \sqrt{I}  =  ( \cos \theta |{\vec x}> |e_x > +\sin \theta |{\vec y}> |e_y
 >),     \label{eqE1}
\end{eqnarray}

the relative amplitudes via the sine and cosine factors allows the
components $|e_x >,  |e_y > $ to be unit normalized. The nonzero
correlation is included  between field components by introducing
the magnitude and phase of the correlation as $<e_x | e_y > \equiv
\alpha $.
 The electric field in Eq. (\ref{eqE1}) can be written as
\begin{eqnarray}
 | \vec{E} >/ \sqrt{I}  =  ( \cos \theta |{\vec x}>  + \alpha \sin \theta |{\vec y}> ) |e_x >
 + \beta \sin \theta |{\vec y}> |{\bar e_x}>,     \label{eqE2}
\end{eqnarray}

 with $|
{\bar e_x }> $ as an orthogonal components of $|e_x >$, such that
$<e_x | {\bar e_x }> =0$ and $ |e_y > = \alpha |e_x > + \beta |{\bar
e_x}>$.

Applying the Schmidt analysis\cite{Eberly2005}  for the state  $|
\vec{E} >/\sqrt{I} $ in Eq. (\ref{eqE2}), the original state can
be written  as:
\begin{eqnarray}
 | \vec{E} > / \sqrt{I} &=& \frac{ \sqrt{\lambda_1}}{
 \sqrt{\eta_1 ^2 + 4 |\alpha|^2  } \sqrt{\zeta_1 ^2 +4|\alpha \beta|^2}} ( \eta_1 |{\vec x}>  + 2 \alpha |{\vec y}> )
  \otimes ( \zeta_1 |e_x > +2 \alpha \beta^*  |{\bar e_x }> )\nonumber \\
      &+&
      \frac{ \sqrt{\lambda_2}}{\sqrt{\eta_2 ^2 + 4|\alpha|^2 }\sqrt{\zeta_2 ^2 +4|\alpha \beta|^2}}
     ( \eta_2 |{\vec x}> + 2 \alpha |{\vec y}> ) \otimes ( \zeta_2 |e_x > + 2 \alpha \beta^*
     |{\bar e_x }> ),     \label{eqSh1}
\end{eqnarray}
where
\begin{eqnarray}
 \eta_1 &=& \cot  \theta - \csc \theta \sec \theta \sqrt{1-|\beta|^2 \sin^2 2 \theta} - \tan \theta\\
  \eta_2 &=&  \cot  \theta + \csc \theta \sec \theta \sqrt{1-|\beta|^2 \sin^2 2 \theta} - \tan \theta\\
 \zeta_1 &=& 1-2 |\beta|^2 + \cot^2 \theta  -\csc^2 \theta \sqrt{1-|\beta|^2 \sin^2
 2 \theta} \\
\zeta_2 &=& 1-2 |\beta|^2 + \cot^2 \theta  + \csc^2 \theta
\sqrt{1-|\beta|^2 \sin^2 2 \theta}. \label{eqetas}
\end{eqnarray}
and the eigenvalues are
\begin{eqnarray}
 \lambda_1 &=& \frac{1}{2} (1-\sqrt{1- |\beta | ^2 \sin^2 2 \theta }) \\
 \lambda_2 &=& \frac{1}{2} (1+\sqrt{1- |\beta | ^2 \sin^2 2 \theta }) .     \label{eigenvalues}
\end{eqnarray}

The Schmidt number as the degree of entanglement can be easily
computed from the eigenvalues $\lambda_1$ and $\lambda_2$
\begin{eqnarray}
 K &=& \frac{1}{\sum_s \lambda_s ^2 }\nonumber \\
 &=&  \frac{1}{1-\frac{1}{2}|\beta|^2 \sin^2 2 \theta}.     \label{Kvalue}
\end{eqnarray}
\begin{figure}
\includegraphics[width=8cm]{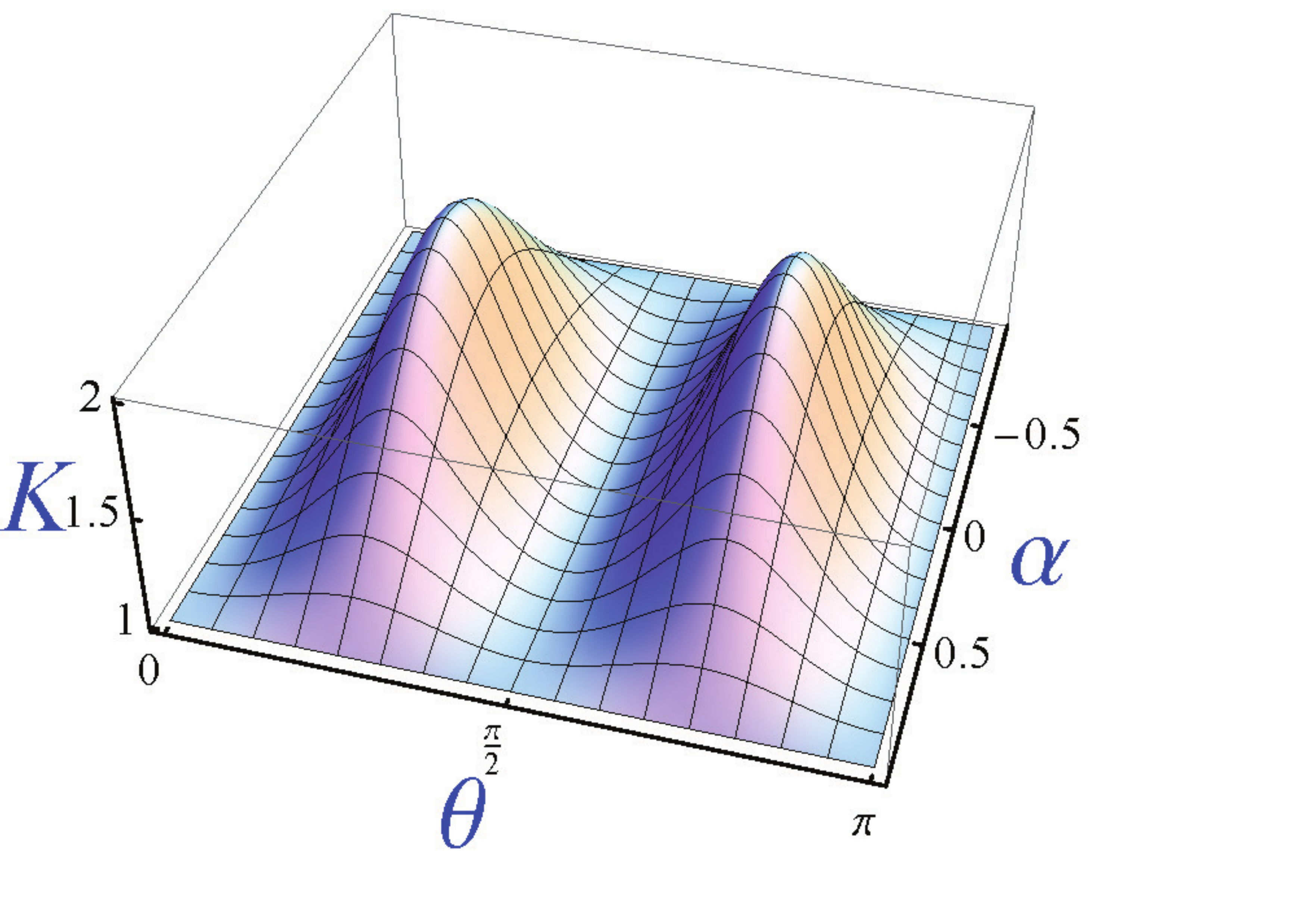}
\caption{Degree of entanglement $K$ with respect to $\alpha$ and
the angle $\theta$.} \label{figk-value}
\end{figure}

We plotted the degree of entanglement $K$ with respect to $\alpha$
and the angle $\theta$ in Fig. \ref{figk-value}.  The degree of
entanglement $K$ may have a maximum  of $2$ under some conditions
such as $\beta = 1$ and $\theta = \frac{\pi}{4}$. Under this
condition, the electric field $E$ in Eq. (\ref{eqSh1}) becomes
\begin{eqnarray}
 | \vec{E} >_{max} / \sqrt{I}  = \frac{1}{\sqrt{2}}
      |{\vec x}>  \otimes |e_x > +
           \frac{1}{\sqrt{2}}
     |{\vec y}>  \otimes |{\bar e_x }> .     \label{eqShmax}
\end{eqnarray}
Mathematically the electric fields $E_x $ and $E_y$ might create
vectors in function space\cite{Eberly2011}, and $ |{\bar e_x }> $
is orthogonal to $|e_x >$. However, it is still unclear how to
make $| \vec{E} >_{max}$ in Eq. \ref{eqShmax}.
 In contrast, it is easy to make a state in which the degree of entanglement is
a minimum value of 1. The degree of entanglement $K$ may have a
minimum of $1$ when $\beta = 0$.  In this case, the electric field
$E$ in Eq. (\ref{eqSh1}) becomes
\begin{eqnarray}
 | \vec{E} >_{m1} / \sqrt{I}  = (\cos \theta |{\vec x}> + \sin \theta |{\vec
 y}>)
   \otimes |e_x > .     \label{eqShmin1}
\end{eqnarray}
This state is simply the linearly polarized state with single
component in function space. We also obtain the minimum $K$ value
$1$ by $\theta =\frac{\pi}{2}$.
\begin{eqnarray}
 | \vec{E} >_{m2}/ \sqrt{I}  =  |{\vec y}> \otimes ( \alpha  |e_x >
 + \beta |{\bar e_x}>),   \label{eqShmin2}
\end{eqnarray}
At this time, the $ | \vec{E} >_{m2}$ state has a single component
in "lab space".

\subsection{Concurrence}

For the electric field $E$ in Eq. (\ref{eqSh1}), we can obtain the
density matrix, $\rho$, based on $ \{ |x \  e_x>, |x \ {\bar
e_x}>, |y \ e_x>, |y \ {\bar e_x }> \}$,

\begin{eqnarray}
\rho_A = \left(\begin{array}{cccc}
\cos^2 \theta & 0 & \alpha^* \cos \theta \sin \theta & \beta^* \cos \theta \sin \theta \\
0             & 0 &             0                    &   0                             \\
 \alpha \cos \theta \sin \theta & 0 & |\alpha|^2 \sin^2 \theta &
 \alpha \beta^* \sin^2 \theta \\
 \beta \cos \theta \sin \theta & 0 & \alpha^* \beta  \sin^2 \theta &
 |\beta|^2 \sin^2 \theta \\
\end{array}\right)   \label{eqrho}
\end{eqnarray}
We can calculate the concurrence of this system from this density
matrix.  The explicit formula for concurrence $C(\rho)$
is:\cite{Wooter1999}
\begin{eqnarray}
 C(\rho) = max \{ 0,\sqrt{\lambda_1} -\sqrt{\lambda_2} -\sqrt{\lambda_3} - \sqrt{\lambda_4 \}},   \label{eqCon}
\end{eqnarray}
where $\lambda_i$ is eigenvalue of $\rho {\bar \rho} $ in
descending order. Here ${\bar \rho}$  is the result of applying
the Pauli operator to $\rho$,
\begin{eqnarray}
{\bar \rho}  = (\sigma_y  \otimes \sigma_y ) \rho^* (\sigma_y
\otimes \sigma_y ). \label{eqRhobar}
\end{eqnarray}
 We obtain the concurrence from the density matrix
 $\rho_A$,
\begin{eqnarray}
C( \rho_A )  = \frac{1}{2} |(\beta +\beta^*) \sin 2 \theta | .
\label{eqCon}
\end{eqnarray}

\begin{figure}
\includegraphics[width=8cm]{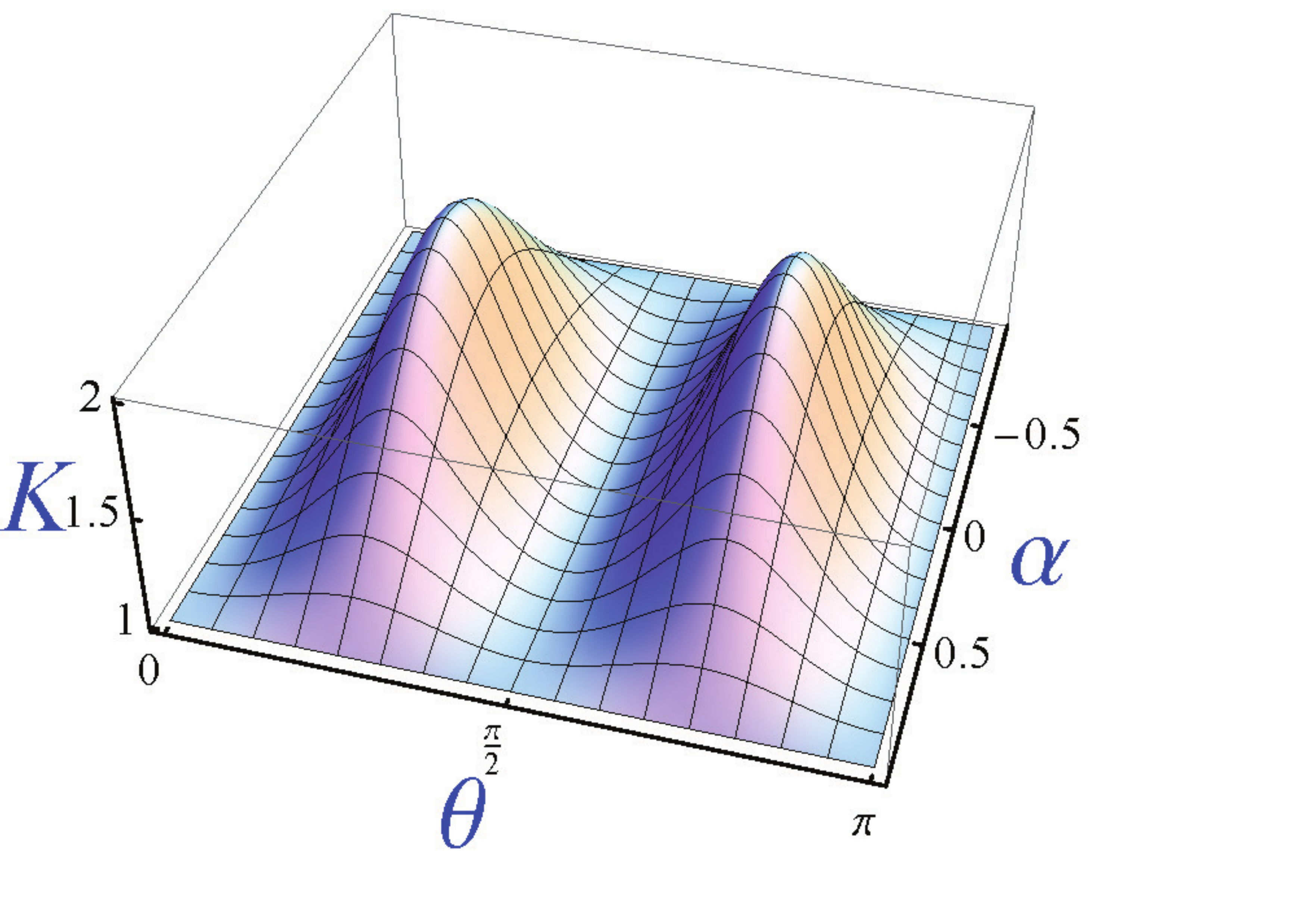}
\caption{The concurrence, $C$, with respect to $\alpha$ and the
angle $\theta$} \label{figc-value}
\end{figure}

We plotted the concurrence $C$ with respect to $\alpha$ and the
angle $\theta$ in Fig. \ref{figc-value}. We plotted the maximum
concurrence for a given $|\beta|$ with $Im(\beta)=0$, as $\beta +
\beta^*$ has its maximum if $\beta$ is a real number.

The concurrence, $C$, is maximum when $ \alpha=0 $ and $\theta =
\frac{\pi}{4}$. Under this condition, the electric field can be
written as in Eq. \ref{eqShmax}. The minimum value of the
concurrence can be obtained with $\theta = \frac {\pi}{2}$ or with
$\beta = 0$ as in Eqs. \ref{eqShmin1} and \ref{eqShmin2}

\section{Conclusion and Discussion}

   Classical light fields may be considered as physical
examples of nonquantum entanglement. Qian and  Eberly reformulated
polarization theory as entanglement analysis. In this perspective,
polarization is a characterization of the correlation between the
vector nature and the statistical nature of the light field. Those
authors discussed the general entanglement Schmidt value K, which
varies from 1 to 3 over the unit polarization sphere for nonplanar
case .
   We applied the concurrence and Schmidt approach to evaluate the degree of
entanglement for a generalized polarization state for the planar-transeverse
case. We found a maximum entanglement state for the two-dimensional
lab space unit vector (${\vec x}$ and ${\vec y}$) and
two-dimensional statistical function space unit vectors ($E_x$ and
$E_y$).
  Although, it is not clear how to measure the degree of
entanglement over lab and  function space, we calculated the
concurrence and the Schmidt value for the usual polarization
states in two-dimensional lab space. We expect some experiments to
measure the degree of entanglement of the maximally entangled
polarization state.

\acknowledgements
 This study was supported by the Basic Science
Research Program, through the National Research Foundation of Korea
(NRF), funded by the Ministry of Education, Science and Technology.

\end{document}